\documentclass[aps,
showpacs,pra,amssymb, amsmath]{revtex4}
\usepackage{amsmath,latexsym,amssymb}
\usepackage[dvips]{graphicx}
\usepackage{amsmath}
\usepackage{latexsym}
\usepackage{amssymb}
\usepackage{bm}

\begin{document}

\title{There is no axiomatic system for the quantum theory}

\author{Koji Nagata}
\affiliation{ Department of Physics, Korea Advanced Institute of
Science and Technology, Daejeon 305-701, Korea}

\date{\today}

\begin{abstract}
Recently, [arXiv:0810.3134] is accepted and published.
We derive an inequality with two settings as tests for the existence of
the Bloch sphere in a spin-1/2 system. 
The probability theory of measurement outcome
within the formalism of von Neumann 
projective measurement violates the inequality.
Namely, we have to give up the existence of the Bloch sphere.
Or, we have to give up the probability theory of measurement outcome 
within the formalism of von Neumann 
projective measurement.
Hence it turns out that there is a contradiction in the Hilbert space formalism of 
the quantum theory, viz.,
there is no axiomatic system for the theory.

\end{abstract}

\maketitle

\section{Introduction}
Recently, \cite{NagataNakamura} is accepted and published.
As a famous physical theory, the quantum theory
(cf. \cite{JJ,Peres,Redhead,Neumann,NIELSEN_CHUANG}) gives accurate and at times remarkably accurate numerical predictions. Much experimental data 
has been agreed with the quantum predictions for long time.

Here we aim to show that 
there is a contradiction in the Hilbert space formalism of 
the quantum theory.
This implies that there is no axiomatic system for the theory.

In what follows, we derive an inequality with two settings as tests for 
the existence of the Bloch sphere, a proposition of the quantum theory, 
in a spin-1/2 system. 
The probability theory of measurement outcome within the formalism of von Neumann projective 
measurement violates the inequality.
What we need is only one spin-1/2 state (a two-dimensional state).
Therefore, there is a contradiction in the set of propositions of 
the quantum theory
in a spin-1/2 system, viz.,
there is no axiomatic system for the quantum theory.

%%%%%%%%%%%%%%%%%%%%%%%%%%

\section{Notation and preparations }\label{Notation}

Throughout this paper, we assume 
von Neumann projective measurement
and
we confine ourselves to the finite-dimensional and 
the discrete spectrum case.
Let ${\bf R}$ denote the reals where $\pm\infty\not\in {\bf R}$.
We assume every eigenvalue in this paper lies in ${\bf R}$.
Further, we assume that every Hermitian operator 
is associated with a unique observable 
because we do not need to distinguish between them in this paper.

We assume the validity of the quantum theory and 
we would like to investigate if
the probability theory of quantum measurement outcome is possible.
Let ${\cal O}$
be the space of Hermitian operators described in a 
finite-dimensional Hilbert
space, and ${\cal T}$ 
be the space of density operators
 described in the Hilbert
space. Namely, ${\cal T}=
\{\psi | \psi\in{\cal O}\wedge\psi\geq 0\wedge {\rm Tr}[\psi]=1\}$.
Now we define the notation $\theta$ which represents 
one result of quantum measurement.
Suppose that 
the measurement of 
a Hermitian operator $A$ for a system in the state $\psi$ yields a 
value $\theta(A)\in {\bf R}$.
Let us consider the following propositions.
Here, $\chi_{\Delta}(x), (x\in{\bf R})$ represents 
the characteristic function.
$\Delta$ is any subset of the reals ${\bf R}$.

{\it Proposition:} BSF ({\it the Born statistical formula}),
\begin{eqnarray}
{\rm Prob}(\Delta)_{\theta(A)}^{\psi}={\rm Tr}[\psi\chi_{\Delta}(A)].
\end{eqnarray}
The whole symbol $(\Delta)_{\theta(A)}^{\psi}$ 
is used to denote the proposition 
that $\theta(A)$ lies in $\Delta$ if the system is in the state $\psi$.
And ${\rm Prob}$ denotes the probability that the proposition holds.

Let us
consider a probability space 
$(\Omega,\Sigma,\mu_{\psi})$, where
$\Omega$ is a nonempty space, $\Sigma$ is a
$\sigma$-algebra of subsets of $\Omega$, and $\mu_{\psi}$ is a
$\sigma$-additive normalized measure on $\Sigma$ such that 
$\mu_{\psi}(\Omega)=1$.
The subscript $\psi$ expresses that
the probability measure is determined uniquely
when the state $\psi$ is specified.

Let us introduce measurable functions 
(random variables) onto $\Omega$
($f: \Omega \mapsto {\bf R}$), which is written as $f_A(\omega)$
for an operator $A\in {\cal O}$.
Here $\omega\in\Omega$.
We introduce appropriate notation.
$P(\omega)\simeq Q(\omega)$ means
$P(\omega)=Q(\omega)$ holds almost everywhere with respect 
to $\mu_{\psi}$ in $\Omega$.

{\it Proposition:} P ({\it the probability theory of 
measurement outcome}),

Measurable function $f_A(\omega)$ exists for 
every Hermitian operator $A$ in ${\cal O}$.

{\it Proposition:} D ({\it the probability distribution rule}),
\begin{eqnarray}
\mu_{\psi}(f^{-1}_{A}(\Delta))={\rm Prob}(\Delta)_{\theta(A)}^{\psi}.
\end{eqnarray}

Now, we see the following:

{\it Lemma:}\cite{Nagata}  Let $S_A$ stand for the 
spectrum of the Hermitian operator $A$.
If
\begin{eqnarray}
{\rm Tr}[\psi A]&=&\sum_{y\in S_A} 
{\rm Prob}(\{y\})_{\theta(A)}^{\psi}y,\nonumber\\
E_{\psi}(A)&:=&\int_{\omega\in \Omega}\mu_{\psi}({\rm d}\omega)
f_{A}(\omega),\nonumber
\end{eqnarray}
then 
\begin{eqnarray}
{\rm P}\wedge{\rm D}
\Rightarrow
{\rm Tr}[\psi A]=E_{\psi}(A).\label{QMHV}
\end{eqnarray}

{\it Proof:}
Note
\begin{eqnarray}
&&\omega\in f^{-1}_{A}(\{y\})\Leftrightarrow
f_{A}(\omega)\in \{y\}\Leftrightarrow
y=f_{A}(\omega),\nonumber\\
&&\int_{\omega\in f^{-1}_{A}(\{y\})}
\frac{\mu_{\psi}({\rm d}\omega)}{\mu_{\psi}
(f^{-1}_{A}(\{y\}))}=1,\nonumber\\
&&y\neq y'\Rightarrow 
f^{-1}_{A}(\{y\})\cap f^{-1}_{A}(\{y'\})=\phi.
\end{eqnarray}
Hence we have
\begin{eqnarray}
&&{\rm Tr}[\psi A]=\sum_{y\in S_A} 
{\rm Prob}(\{y\})_{\theta(A)}^{\psi}y
=\sum_{y\in{\bf R}} 
{\rm Prob}(\{y\})_{\theta(A)}^{\psi}y
=\sum_{y\in{\bf R}}\mu_{\psi}(f^{-1}_{A}(\{y\}))y\nonumber\\
&&=\sum_{y\in{\bf R}}
\mu_{\psi}(f^{-1}_{A}(\{y\}))y
\times \int_{\omega\in f^{-1}_{A}(\{y\})}
\frac{\mu_{\psi}({\rm d}\omega)}{\mu_{\psi}
(f^{-1}_{A}(\{y\}))}\nonumber\\
&&=\sum_{y\in{\bf R}}\int_{\omega\in f^{-1}_{A}(\{y\})}
\mu_{\psi}(f^{-1}_{A}(\{y\}))
\times 
\frac{\mu_{\psi}({\rm d}\omega)}{\mu_{\psi}
(f^{-1}_{A}(\{y\}))}f_{A}(\omega)\nonumber\\
&&=\int_{\omega\in \Omega}\mu_{\psi}({\rm d}\omega)
f_{A}(\omega)=E_{\psi}(A).
\end{eqnarray}
QED.

Thus, one may
assume the probability measure $\mu_{\psi}$ is 
chosen such that the following relation
is valid:
\begin{eqnarray}
{\rm Tr}[\psi A]=\int_{\omega\in\Omega}\mu_{\psi}({\rm d}\omega)
f_A(\omega)
\end{eqnarray}
for every Hermitian operator $A$ in ${\cal O}$.

From BSF, P, and D, 
the possible value of $f_A(\omega)$ takes eigenvalues of $A$ 
almost everywhere with respect 
to $\mu_{\psi}$ in $\Omega$.
That is, we have the following theorem.

{\bf Theorem:} ({\it The possible values of measurement outcome}):

Let $S_A$ stand for the spectrum of the Hermitian operator $A$.
For every quantum state described in a Hilbert 
space ${\cal H}$, 
\begin{eqnarray}
{\rm BSF}\wedge{\rm P}\wedge
{\rm D}
\Rightarrow
f_A(\omega)\in S_A,~(\mu_{\psi}-a.e.).\label{MO}
\end{eqnarray}

\section{There is no axiomatic system for the quantum theory}

Here, we shall show the following theorem.

{\bf Theorem:}
\begin{eqnarray}
[{\rm The\ Bloch\ sphere\ exists.}]\wedge{\rm BSF}\wedge{\rm P}\wedge
{\rm D}
\Rightarrow
\bot.
\end{eqnarray}
{\it Proof:}
Assume a spin-$1/2$ state $\psi$.
Let $\vec \sigma$ be $(\sigma_x,\sigma_y,\sigma_z)$, 
the vector of Pauli operator.
The measurements (observables) on a spin-1/2 state 
of $\vec n\cdot\vec\sigma$ are parameterized by 
a unit vector $\vec n$ (direction along which the spin component is measured).
Here, $\cdot$ is the scalar product in ${\bf R}^{\rm 3}$.

One has a quantum expectation value $E_{\rm QM}$ as
\begin{eqnarray}
E_{\rm QM}\equiv {\rm Tr}[\psi \vec n_k\cdot \vec \sigma],~k=1,2.\label{et}
\end{eqnarray}
One has $\vec x\equiv x^{(1)}$, $\vec y\equiv x^{(2)}$, and 
$\vec z\equiv x^{(3)}$ which are the Cartesian axes relative to which spherical angles are measured.
Let us write the two unit vectors in the 
plane defined by $x^{(1)}$ and $x^{(2)}$ in the following way:
\begin{eqnarray}
\vec n_k=\sin\theta_k \vec x^{(1)}+\cos\theta_k \vec x^{(2)}.
\label{vector}
\end{eqnarray}
Here, the angle $\theta_k$ takes two values:
\begin{eqnarray}
\theta_1=0,~\theta_2=\frac{\pi}{2}.
\end{eqnarray}

We shall derive a necessary condition for
the quantum expectation value
for the system in a spin-1/2 state given in (\ref{et}).
Namely, we shall derive the value of the scalar product 
$(E_{\rm QM}, E_{\rm QM})$ 
of the quantum expectation value, 
$E_{\rm QM}$ given in (\ref{et}).
We use decomposition (\ref{vector}).
We introduce simplified notations as
\begin{eqnarray}
T_{i}=
{\rm Tr}[\psi \vec{x}^{(i)}\cdot \vec \sigma ]
\end{eqnarray}
and
\begin{eqnarray}
(c^1_k, c^2_k,)=(\sin \theta_k,
\cos\theta_k).
\end{eqnarray}
Then, we have
\begin{eqnarray}
&&(E_{\rm QM}, E_{\rm QM})  = 
\sum_{k=1}^2
\left(\sum_{i=1}^2T_{i}
c^{i}_k\right)^2  =  
\sum_{i=1}^2T_{i}^2\leq 1,
\label{EEvalue}
\end{eqnarray}
where we have used the orthogonality relation
$\sum_{k=1}^2 ~ c_k^{\alpha} c_k^{\beta}  =  \delta_{\alpha,\beta}$.
From a proposition of the quantum theory, the Bloch sphere, the value of 
$\sum_{i=1}^2T_{i}^2$ is bounded as 
$\sum_{i=1}^2T_{i}^2\leq 1$.
Clearly, the reason of the condition 
(\ref{EEvalue}) is the Bloch sphere
\begin{eqnarray}
\sum_{i=1}^3 
({\rm Tr}[\psi \vec{x}^{(i)}\cdot \vec \sigma])^2\leq 1.
\end{eqnarray}
Thus a violation of the inequality (\ref{EEvalue})
implies a violation of the existence of the Bloch sphere 
(in a spin-1/2 system).

Let us assume BSF, P, and D hold. In this case, a quantum expectation value, which is the average of the results of 
projective measurements (based on a probability space) is given by
\begin{equation}
E_{\rm QM}=\int_{\omega\in\Omega}\mu_{\psi}({\rm d}\omega)
f_{\vec n}(\omega).\label{avg}
\end{equation}
The possible values of $f_{\vec n}(\omega)$ are $\pm 1$ (in $\hbar/2$ unit) almost everywhere with respect 
to $\mu_{\psi}$ in $\Omega$.

We shall derive a necessary condition for 
the quantum expectation value obtained by the projective measurements
on a spin-1/2 state given in (\ref{avg}).
We shall derive the absolute value of 
the scalar product $|(E_{\rm QM}, E_{\rm QM})|$
of the quantum expectation value, 
$E_{\rm QM}$ given in (\ref{avg}).
One has
\begin{eqnarray}
|(E_{\rm QM}, E_{\rm QM})|&=&\bigg|\sum_{k=1}^2
\left(\int_{\omega\in\Omega}\mu_{\psi}({\rm d}\omega)f_{\vec n_k}(\omega)
 \times 
 \int_{\omega'\in\Omega}\mu_{\psi}({\rm d}\omega')f_{\vec n_k}(\omega')
 \right)
\bigg|\nonumber\\
&=&\bigg|\sum_{k=1}^2
\left(\int_{\omega\in\Omega}\mu_{\psi}({\rm d}\omega)
\int_{\omega'\in\Omega}\mu_{\psi}({\rm d}\omega')
f_{\vec n_k}(\omega)
f_{\vec n_k}(\omega')
 \right)
\bigg|\nonumber\\
&\simeq&\bigg|\sum_{k=1}^2
\left(\int_{\omega\in\Omega}\mu_{\psi}({\rm d}\omega)
\int_{\omega'\in\Omega}\mu_{\psi}({\rm d}\omega')
 \right)
\bigg|
 =2.
\label{integral}
\end{eqnarray}
We have used the following, viz., 
 \begin{eqnarray}
{\rm BSF}\wedge{\rm P}\wedge
{\rm D}
\Rightarrow
f_{\vec n_k}(\omega)\in \{\pm 1\},~(\mu_{\psi}-a.e.).
\end{eqnarray}
Therefore, one has the value (\ref{integral}) in contradiction to 
(\ref{EEvalue}).
%%%%%%%%%%%%%%%%%%%%%%%%%%%%%%%%%%%%%%%%%%%%%%%%%%%%%%%%
As we have shown, it should be that $(E_{\rm QM}, E_{\rm QM})\leq 1$ if we accept the existence of the Bloch sphere.
However, using BSF, P, and D, the probability theory of the results of von Neumann projective measurements violates the inequality since $|(E_{\rm QM}, E_{\rm QM})|=2$.
Namely, we have to give up, at least, one of propositions,
the Bloch sphere, BSF, P, and D.
QED.

\section{Summary}

In summary, the probability theory of the results of von Neumann projective measurements cannot allow the existence of the Bloch sphere. These quantum-theoretical propositions must contradict each other.
Therefore there is a contradiction in the set of propositions of the quantum theory.
Hence there is no axiomatic system for the quantum theory.
Our result was obtained in a quantum system which is in a spin-1/2 state.

%%%%%%%%%%%%%%%%%%%%%%%%%%%%%%%%%%%%

\acknowledgments
This work has been
supported by Frontier Basic Research Programs at KAIST and K.N. is
supported by the BK21 research professorship.

\end{document}